\documentclass[prl,twocolumn,superscriptaddress,showpacs]{revtex4-1}

\usepackage[T1]{fontenc}
\usepackage{color,graphicx}
\usepackage{times}
\usepackage[dvipdfmx,colorlinks=true]{hyperref}

\usepackage{amsmath,amsthm,amssymb}

\newcommand\ket[1]{\ensuremath{|#1\rangle}}
\newcommand\bra[1]{\ensuremath{\langle#1|}}

\newcommand\prsect[1]{\vspace{.3em}\noindent{\em #1.---\/}}

\begin{document}


\title{Correlations in excited states of local Hamiltonians}

\author{Jianxin Chen}%
\affiliation{Department of Mathematics \& Statistics, University of
  Guelph, Guelph, Ontario, Canada}%
\affiliation{Institute for Quantum Computing and School of Computer
  Science, University of Waterloo, Waterloo, Ontario, Canada}%
\author{Zhengfeng Ji}%
\affiliation{Institute for Quantum Computing and School of Computer
  Science, University of Waterloo, Waterloo, Ontario, Canada}%
\affiliation{State Key Laboratory of Computer Science, Institute of
  Software, Chinese Academy of Sciences, Beijing, China}%
\author{Zhaohui Wei}%
\affiliation{Centre for Quantum Technologies, National University of
  Singapore, Singapore 117543, Singapore}%
\author{Bei Zeng}%
\affiliation{Department of Mathematics \& Statistics, University of
  Guelph, Guelph, Ontario, Canada}%
\affiliation{Institute for Quantum Computing, University of Waterloo,
  Waterloo, Ontario, Canada}%

\begin{abstract}
  Physical properties of the ground and excited states of a $k$-local
  Hamiltonian are largely determined by the $k$-particle reduced
  density matrices ($k$-RDMs), or simply the $k$-matrix for fermionic
  systems---they are at least enough for the calculation of the ground
  state and excited state energies. Moreover, for a non-degenerate
  ground state of a $k$-local Hamiltonian, even the state itself is
  completely determined by its $k$-RDMs, and therefore contains no
  genuine ${>}k$-particle correlations, as they can be inferred from
  $k$-particle correlation functions. It is natural to ask whether a
  similar result holds for non-degenerate excited states. In fact, for
  fermionic systems, it has been conjectured that any non-degenerate
  excited state of a $2$-local Hamiltonian is simultaneously a unique
  ground state of another $2$-local Hamiltonian, hence is uniquely
  determined by its $2$-matrix. And a weaker version of this
  conjecture states that any non-degenerate excited state of a
  $2$-local Hamiltonian is uniquely determined by its $2$-matrix among
  all the pure $n$-particle states. We construct explicit
  counterexamples to show that both conjectures are false. It means
  that correlations in excited states of local Hamiltonians could be
  dramatically different from those in ground states. We further show
  that any non-degenerate excited state of a $k$-local Hamiltonian is
  a unique ground state of another $2k$-local Hamiltonian, hence is
  uniquely determined by its $2k$-RDMs (or $2k$-matrix).
\end{abstract}

\date{\today}


\maketitle

In many-body quantum systems, correlations in quantum states, both
ground states and excited states, play an important role for many
interesting physics phenomena, ranging from high temperature
superconductivity, fractional quantum Hall effect to various kind of
quantum phase transitions. Traditionally, correlation is characterized
by correlation functions of local physical observables. To better
understand the structure of many-body correlations, however, we need a
method to separate out the contribution of the amount that comes from
essentially fewer-body correlations. Irreducible $k$-party
correlation~\cite{LPW02,Zho08}, a concept originating from information
theoretical ideas, provides such a method to quantify many-body
correlations. Especially, an $n$-particle pure state $\ket{\psi}$
contains no irreducible ${>}k$-party correlation if it is uniquely
determined by its $k$-particle reduced density matrices ($k$-RDMs),
meaning that there does not exist any other $n$-particle state, pure
or mixed, which has the same $k$-RDMs as those of $\ket{\psi}$.

As a physical interpretation of irreducible correlations, we note that
the non-degenerate ground state of a $k$-local Hamiltonian contains no
irreducible ${>}k$-party correlation. More concretely,
the Hamiltonian $H$ of a real $n$-particle system usually involves
terms of at most $k$-body interactions, where $k$ is a small
number~\cite{Has10}. This kind of Hamiltonian is called $k$-local and
for most physical systems $k=2$. If $\ket{\psi_0}$ is a ground state
of $H$, then the ground state energy $E_0 = \bra{\psi_0}H\ket{\psi_0}$
is determined by the $k$-RDMs of $\ket{\psi_0}$. Generically, the
ground state will be non-degenerate. In this case, $\ket{\psi_0}$ is
uniquely determined by its $k$-RDMs, because if there exists any other
$n$-particle state, pure or mixed, which has the same $k$-RDMs as
those of $\ket{\psi_0}$, then there must be another pure state which
has the same energy as $\ket{\psi_0}$, making the ground space
degenerate. This kind of ``unique determination'' legitimates, in a
very strong sense, the reduced density matrix approach for many-body
systems (cf.~Ref.~\cite{CY00}).

Similar observation applies for fermionic systems, namely, the unique
ground state of a $k$-local fermionic Hamiltonian is uniquely
determined by its $k$-matrix. Indeed, related studies for fermionic
systems in quantum chemistry date back to early
1960s~\cite{Cio00,CY00}, where the properties of both ground states
and excited states of $2$-local fermionic Hamiltonians were studied.
For excited states, it is conjectured that any non-degenerate excited
state of a $2$-local fermionic Hamiltonian is simultaneously a unique
ground state of another $2$-local fermionic Hamiltonian, hence is
uniquely determined by its $2$-matrix~\cite{Cio00}. And a weaker
version of this conjecture states that any non-degenerate excited
state of a $2$-local fermionic Hamiltonian is uniquely determined by
its $2$-matrix among all the {\it pure} $n$-particle fermionic
states~\cite{Maz98}.

If these conjectures were true, then understanding the excited-state
properties of a system of $N$-fermions could be restricted in studying
the set of all the $2$-matrices, whose characterization is called the
$N$-representability problem in quantum chemistry~\cite{CY00}. The
$N$-representability problem has been studied extensively in the past
several decades and significant progresses in studying practical
chemical systems have been made~\cite{CY00,Cio00,Maz98}, though this
problem is shown to be difficult in the most general
settings~\cite{LCV07}. Meanwhile, it is also natural to ask whether
similar conjectures may hold for excited states of $k$-local spin
Hamiltonians, as excited states are also important for characterizing
nice physics phenomena, especially in non-zero temperature situation.
Sometimes even the zero temperature physics cannot be characterized by
ground states only, for instance, in certain kind of quantum phase
transitions~\cite{TL03}.

Here we construct explicit counterexamples to show that both
conjectures for fermionic systems are false. In more general settings
of $n$-particle systems, not necessarily fermionic, we further show
that any non-degenerate excited state of a $k$-local Hamiltonian is a
unique ground state of another $2k$-local Hamiltonian, hence is
uniquely determined by its $2k$-RDMs. This implies, for fermionic
systems with $2$-local Hamiltonians, that the understanding of some
properties of excited states will need the information of their
$4$-matrices. We also apply our understanding to the study of
correlations in $n$-qubit symmetric Dicke states~\cite{Dic54} and show
that they are uniquely determined by their $2$-RDMs. We believe that
our result sets a good starting point for studying excited-state
properties of many-body systems based on the reduced density matrix
approach, and will lead to fruitful results in related areas,
including quantum information science, quantum chemistry and many-body
physics.

\prsect{From spin systems to fermion systems} To relate the fermionic
problem to known results in quantum information theory, we need a map
from a qubit system to a fermionic system. We now show how to map an
$n$-qubit system to a fermionic system, with $N=n$ fermions and $M=2N$
modes. This map has been already discussed in~\cite{LCV07,OCZ+11}, so
we briefly review the construction here. The idea is to represent each
qubit $s$ as a single fermion that can be in two different modes
$a_i,b_i$, so each $n$-qubit basis state corresponds to the following
$N$-fermion state:
\begin{equation*}
  \ket{z_1,\ldots,z_n} \mapsto
  (a_1^\dagger)^{1-z_1} (b_1^\dagger)^{z_1} \cdots
  (a_n^\dagger)^{1-z_n} (b_n^\dagger)^{z_n} \ket{\Omega},
\end{equation*}
where $z_i=0,1$ and $\ket{\Omega}$ is the vacuum state. Also, all the
relevant single-qubit Pauli operators can be mapped via
\begin{equation*}
  X_i \mapsto a_i^\dagger b_i + b_i^\dagger a_i,\;
  Y_i \mapsto i\left( b_i^\dagger a_i - a_i^\dagger b_i \right),\;
  Z_i \mapsto \openone - 2 b_i^\dagger b_i.
\end{equation*}
In addition, one needs to add the following projectors as extra terms
in the fermionic Hamiltonian:
\begin{equation*}
  P_i=(2a_i^\dagger a_i-\openone)(2b_i^\dagger b_i-\openone).
\end{equation*}
As all the $P_i$'s are biquadratic and commute with all the
single-qubit Pauli operators, the complete Hamiltonian will be block
diagonal. By making the weights of these projectors large enough, we
can always guarantee that the ground state of the full Hamiltonian
will have exactly one fermion per site.

Therefore, to disprove both conjectures for fermionic systems, one
only needs to find counterexamples in $n$-qubit systems. In other
words, we will need to find an $n$-qubit pure state $\ket{\psi}$ which
is a non-degenerate eigenstate of some $2$-local Hamiltonian, but
there exists another pure state $\ket{\psi}'$ which has the same
$2$-RDMs as those of $\ket{\psi}$. Therefore, $\ket{\psi}$ cannot be a
unique ground state of any $2$-local Hamiltonian. Then by applying the
spin-to-fermion map discussed above, one can result in a
counterexample for the fermionic case.

\prsect{Simple counterexamples from $3$-qubit states} To construct
explicit counterexamples, we start from the simplest case of $n=3$.
First of all, we need a state $\ket{\psi}$ which is not uniquely
determined by its $2$-RDMs and then further show that $\ket{\psi}$ is
a non-degenerate eigenstate of some $2$-local Hamiltonian $H=\sum_i
H_i$, where each $H_i$ acts non-trivially on at most two qubits. It is
well known that almost all $3$-qubit states are uniquely determined by
their $2$-RDMs except those locally equivalent to the GHZ-type states
$\alpha\ket{000}+\beta\ket{111}$, for $\alpha,\beta\neq
0$~\cite{LPW02,GHZ89}. Up to local unitary operations, one only needs
to consider the case where $\alpha,\beta$ are real. Apparently, the
pure state $\alpha\ket{000}+\beta e^{i\theta}\ket{111}$ has the same
$2$-RDMs as those of $\alpha\ket{000}+\beta\ket{111}$, so
$\alpha\ket{000}+\beta\ket{111}$ is not uniquely determined by its
$2$-RDMs, even among pure states.

To show that $\alpha\ket{000}+\beta\ket{111}$ can be a non-degenerate
eigenstate of some $2$-local Hamiltonian, we construct the $2$-local
Hamiltonian explicitly. We start from a simple case of the GHZ state
where $\alpha = \beta = 1/\sqrt{2}$, $\ket{\psi}_{\text{GHZ}}
= (\ket{000} + \ket{111})/\sqrt{2}$. The GHZ state is the
eigenstate of the commuting Pauli operators $Z_1Z_2, Z_2Z_3,
X_1X_2X_3$ with eigenvalue $1$, where $X_i, Y_i, Z_i$ stands for the
Pauli $X, Y, Z$ operators acting on the $i$-th qubit. In the language
of stabilizers~\cite{Got97}, $\ket{\psi}_{\text{GHZ}}$ is stabilized
by the group generated by $Z_1Z_2, Z_2Z_3, X_1X_2X_3$.

For the Hamiltonian $H_0 = - Z_1Z_2 - Z_2Z_3$,
$\ket{\psi}_{\text{GHZ}}$ is an eigenstate but degenerate with any
state in the space spanned by $\ket{000}, \ket{111}$. In order to
remove the $2$-fold degeneracy and to make $\ket{\psi}_{\text{GHZ}}$ a
non-degenerate eigenstate, we note that $X_1X_2X_3
\ket{\psi}_{\text{GHZ}} = \ket{\psi}_{\text{GHZ}}$. Therefore,
$\ket{\psi}_{\text{GHZ}}$ is an eigenstate of $H_1 = X_1X_2 - X_3$
with eigenvalue $0$, which is not the case for any other state in the
space spanned by $\ket{000},\ket{111}$. Finally, one concludes that
$\ket{\psi}_{\text{GHZ}}$ is a non-degenerate eigenstate of the
$2$-local Hamiltonian $H = -Z_1Z_2-Z_2Z_3 + c(X_1X_2-X_3),$ for a
properly chosen $c$ (for instance, one can choose $c=-1$ then
$\ket{\psi}_{\text{GHZ}}$ is the non-degenerate first excited state of
$H$, with energy $-2$.)

For the state $\alpha\ket{000}+\beta\ket{111}$, similar ideas apply.
Denote a $2\times 2$ diagonal matrix with diagonal elements $a_{11},
a_{22}$ by $\text{diag}(a_{11},a_{22})$, then we have
\begin{equation*}
  \text{diag}(\frac{\beta}{\alpha}, \frac{\alpha}{\beta})_1 X_1X_2X_3
  (\alpha\ket{000}+\beta\ket{111}) = \alpha\ket{000}+\beta\ket{111},
\end{equation*}
where the operator $\text{diag}(\frac{\beta}{\alpha},
\frac{\alpha}{\beta})_i$ acts on the $i$-th qubit. Therefore,
$\alpha\ket{000} + \beta\ket{111}$ is a non-degenerate eigenstate of
the $2$-local Hamiltonian
\begin{equation*}
  H= a Z_1Z_2 + b Z_2Z_3 + c\, (\text{diag}(\frac{\beta}{\alpha},
  \frac{\alpha}{\beta})_1 X_1X_2-X_3),
\end{equation*}
for some properly chosen $a,b,c$.

These $3$-qubit examples can thus be mapped to fermionic
counterexamples of three fermions with six modes, thus disprove the
conjecture discussed in~\cite{Cio00} and its weaker version
in~\cite{Maz98}.

\prsect{More counterexamples} One may think that the existence of the
counterexamples from $3$-qubit states is due to the fact that almost
all (except the GHZ-type) $3$-qubit states are uniquely determined by
their $2$-RDMs, and hope that these conjectures could actually hold
for most of the other cases. Here we show that the above discussion of
the counterexamples from $3$-qubit states provides a systematic way to
find a large class of counterexamples.

The idea of constructing the counterexamples from $3$-qubit states is
the following: start from a $2$-local Hamiltonian $H_0$ whose ground
space is degenerate (for simplicity, we assume it is two-fold
degenerate). Choose a basis $\ket{C_0}$ and $\ket{C_1}$ for the ground
space of $H_0$ such that: 1) $\ket{C_0}$ and $\ket{C_1}$ have the same
$2$-RDMs; 2) there exists a weight $3$ or $4$ operator $M$ such that
$M\ket{C_0}=\ket{C_0}$ but $M\ket{C_1}\neq \ket{C_1}$. Then one can
`decompose' the operator $M$ into a $2$-local one $H_1$ such that
$\ket{C_0}$ is an eigenvector with eigenvalue zero (for instance, if
$M=X_1X_2Z_3Z_4$, one can chose $H_1=X_1X_2-Z_3Z_4$), then the
Hamiltonian $H=H_0+cH_1$ will have $\ket{C_0}$ as a non-degenerate
eigenstate for a properly chosen $c$. Thus $\ket{C_0}$ gives a
counterexample after applying the spin-to-fermion map.

In general, for a given $H_0$ one cannot guarantee the existence of
such $\ket{C_0}$, $\ket{C_1}$ and $M$. However, in certain case of
quantum error-correcting codes~\cite{Got97}, they are easy to find.
Consider a quantum error-correcting code of dimension $>1$ which is a
ground state of a $2$-local Hamiltonian, with distance $3$ or $4$.
Then any state in the code space has the same
$2$-RDMs~\cite{OCZ+11,BH10}, so one can easily find $\ket{C_0}$ and
$\ket{C_1}$ which are orthogonal. If the code is a stabilizer or
stabilizer subsystem code, then the logical operator $M$ which
satisfies $M\ket{C_0}=\ket{C_0}$ and $M\ket{C_1}=-\ket{C_1}$ will be a
Pauli operator of weight $3$ (if the code distance is $3$) or $4$ (if
the code distance is $4$).

One simple example is the Bacon-Shor code on a $3\times 3$ (or
$4\times 4$) square lattice~\cite{Bac06}. We discuss the $3\times 3$
case for simplicity. The system consists of $n=9$ qubits arranged on a
$3\times 3$ square lattice, and the Hamiltonian is given by
\begin{equation*}
  H_0 = -J_x\sum_{j,k}X_{j,k}X_{j+1,k} - J_z\sum_{k,j}Z_{j,k}Z_{j,k+1},
\end{equation*}
whose ground space is two-fold degenerate, constituting a quantum
error-correcting code of distance $3$, where $J_x, J_z>0$, the
subscripts $j,k$ refer to the qubit of the $j$-th row and $k$-th
column and the addition is $\text{mod} 3$. An orthonormal basis of the
code space can be chosen as $\ket{C_0}$ and $\ket{C_1}$ such that the
logical $Z$ operator $\bar{Z}$, satisfying $\bar{Z} \ket{C_0} =
\ket{C_0}$ and $\bar{Z} \ket{C_1} = -\ket{C_1}$, is given by $\bar{Z}
= Z_{1,1} Z_{2,1} Z_{3,1}$~\cite{Bac06}. Therefore, $\ket{C_0}$ is a
non-degenerate eigenstate of the $2$-local Hamiltonian
\begin{equation*}
  H = H_0 + c\, (Z_{1,1}Z_{2,1}-Z_{3,1})
\end{equation*}
for a properly chosen $c$.

\prsect{Correlations in excited states} We would like to consider this
problem in more general settings of $n$-particle states which are not
necessarily fermionic, or do not have any kind of symmetry. Our method
directly generalizes to the case of $k>2$, showing that a
non-degenerate eigenstate of a $k$-local Hamiltonian may not be
uniquely determined by its $k$-RDMs, even among pure states. A simple
example could be the $n$-particle GHZ state
\begin{equation*}
  \ket{\psi^{(n)}}_{\text{GHZ}} = \frac{1}{\sqrt{2}}
  (\ket{0}^{\otimes n} + \ket{1}^{\otimes n}).
\end{equation*}

For simplicity we take $n$ even (the odd cases can be dealt with
similarly). Note the GHZ state is not uniquely determined by its
$(n-1)$-RDMs, as the state $\frac{1}{\sqrt{2}} (\ket{0}^{\otimes n} +
e^{i\theta}\ket{1}^{\otimes n})$ has the same $(n-1)$-RDMs.

Using similar ideas of the $3$-qubit case, we know that
$\ket{\psi^{(n)}}_{\text{GHZ}}$ is a non-degenerate ground state of the
$\frac{n}{2}$-local Hamiltonian
\begin{equation*}
  \begin{split}
    & H = -Z_1Z_2 -Z_2Z_3 -\cdots -Z_{n-1}Z_n\\
    & + c\, (X_1X_2\cdots X_{n/2} - X_{n/2+1} X_{n/2+2} \cdots X_n),
  \end{split}
\end{equation*}
for a properly chosen $c$. Using the idea based on quantum
error-correcting codes, one can also find other states which are
non-degenerate eigenstates of a $k$-local Hamiltonian but are not
uniquely determined by their $k$-RDMs, even among pure states.

Given that a unique ground state of a $k$-local Hamiltonian is
uniquely determined by its $k$-RDMs, these examples show that
correlations in excited states of local Hamiltonians could be
dramatically different from correlations in the ground states. Then an
interesting question arises: how `dramatic' this correlation could be
for non-degenerate eigenstates of local Hamiltonians? More concretely,
can a non-degenerate eigenstate of a $k$-local Hamiltonian have
non-zero irreducible $r$-party correlations for any $r\leq n$? This
question becomes more intriguing when $k$ is a constant independent of
$n$. That is, can a non-degenerate eigenstate of a local Hamiltonian
have non-local irreducible correlations?

We show, however, this is not the case---a non-degenerate eigenstate
of a $k$-local Hamiltonian is uniquely determined by its $2k$-RDMs and
, therefore, cannot have ${>}2k$-party irreducible correlation. To see
this, let us consider a non-degenerate eigenstate $\ket{\psi}$ of a
$k$-local Hamiltonian $H$ with $H\ket{\psi} = h\ket{\psi}$, and
without loss of generality, assume $h=0$. Then, $H^2\ket{\psi}=0$, and
$\ket{\psi}$ becomes the ground state of $H^2$. Because $H$ is
$k$-local, $H^2$ is at most $2k$-local, $\ket{\psi}$ is then uniquely
determined by its $2k$-RDMs. This result shows that although
correlations in non-degenerate excited states of a local Hamiltonians
are different from those in ground states, they are still `local'
irreducible correlations.

We mention that the $2k$ bound is tight, as there exists
a non-degenerate excited state of a $k$-local Hamiltonian that is not
uniquely determined by its $(2k{-}1)$-RDMs. One simple example is the
GHZ state of $2k$ qubits, which is a non-degenerate excited state of a
$k$-local Hamiltonian, but is not uniquely determined by its
$(2k{-}1)$-RDMs.

It is also easy to see that the discussion here about non-degenerate
eigenstates can be directly extended to the degenerate case. That is,
if $V$ is an eigenspace of a $k$-local Hamiltonian, then $V$ is a
ground space of a $2k$-local Hamiltonian.

\prsect{Applications} We have discussed in a very general setting
about correlations in excited states of local Hamiltonians. It turns
out that our techniques can also help to understand correlations in
certain quantum states in a relatively simple way. As an example, we
discuss correlations in $n$-qubit symmetric Dicke states.

The $n$-qubit symmetric Dicke state $\ket{W_n(i)}$ ($i=0,1,\ldots, n$)
is the equal weight superposition of weight-$i$ bit
strings~\cite{Dic54}. For instance, $\ket{W_n(0)}=\ket{00\cdots 0}$,
and $\ket{W_n(1)} = (\ket{10\cdots 0} + \ket{01\cdots 0}
+\cdots +\ket{00\cdots 1})/\sqrt{n}$ is the $n$-qubit $W$ state.

As $\ket{W_n(0)}$ and $\ket{W_n(n)}$ are product states, they are
uniquely determined by their $1$-RDMs. We know that $\ket{W_n(1)}$ is
uniquely determined by its $2$-RDMs~\cite{PR08,*PR09}, and the case
for $\ket{W_n(i)}$ ($i=2,3,\ldots, n-2$) remain open. Here we show
that $\ket{W_n(i)}$ is uniquely determined by its $2$-RDMs for any
$i$. Note, however, that non-symmetric Dicke states, which are
non-equal weight superposition of weight-$i$ bit strings, are in
general not uniquely determined by their $2$-RDMs~\cite{PR08,*PR09}.

To begin with, we define a collective operator $S_z=\sum_{i=j}^n Z_j$,
the $Z$ component of the total angular momentum of the system.
Obviously for a given $i$, $\ket{W_n(i)}$ is an eigenstate of $S_z$,
which is in general degenerate. For a properly chosen constant $c_i$,
$\ket{W_n(i)}$ could be an eigenvalue zero eigenstate of
$H_0=S_z+c_i\openone$.

For a given $i$, $\ket{W_n(i)}$ is then the ground state of the
$2$-local Hamiltonian $H_0^2$. The ground space is in general
degenerate, however, $\ket{W_n(i)}$ is the only state in the ground
space which is invariant under the permutation of any two qubits. To
split the degeneracy and to make $\ket{W_n(i)}$ the unique ground
state, note the two-qubit SWAP operator $\text{SWAP}_{jk}\ket{x}_j
\ket{y}_k = \ket{y}_j\ket{x}_k$ has eigenvalues $1$ and $-1$. For any
$j,k$, $\text{SWAP}_{jk} \ket{W_n(i)} = \ket{W_n(i)}$. Therefore,
$\ket{W_n(i)}$ is the unique ground state of the $2$-local
Hamiltonian,
\begin{equation*}
  H=H_0^2-c\sum_{j<k}\text{SWAP}_{jk}
\end{equation*}
for small enough $c>0$, hence $\ket{W_n(i)}$ is uniquely determined by
its $2$-RDMs.

\prsect{Conclusion} We have discussed the correlations in excited
states of local Hamiltonians. Explicit examples are constructed to
show that, a non-degenerate excited state of a $k$-local Hamiltonian
may not be uniquely determined by its $k$-RDMs, even among pure
states. By applying a spin-to-fermion map, these examples disprove a
conjecture in quantum chemistry, as well as a weaker version,
regarding non-degenerate excited states of $2$-local Hamiltonians in
fermionic systems. Therefore, to understand the properties of the
excited states of a $2$-local fermionic system, the information in
$2$-matrices may not be enough and one has to resort to $4$-matrices
in some cases.

We further showed that any non-degenerate excited state of a $k$-local
Hamiltonian is a unique ground state of another $2k$-local
Hamiltonian, hence is uniquely determined by its $2k$-RDMs. Moreover,
this $2k$ bound is indeed optimal. For a constant $k$, this result
indicates that a non-degenerate excited state cannot have `non-local'
irreducible correlations.

Our techniques also helped us to understand correlations in certain
quantum states in a relatively simple way. As an example, we have
shown that all the $n$-qubit symmetric Dicke states are uniquely
determined by their $2$-RDMs.

In conclusion, our work corrects some misconceptions about the
excited states of $k$-local Hamiltonians and provides the basis for
further investigation of excited state properties of many-body quantum
systems. We hope that our investigations will help to build new
connections between quantum information science, quantum chemistry and
many-body physics.

\prsect{Acknowledgement} JC is supported by NSERC. ZJ acknowledges
support by NSERC and NSF of China (Grant No. 60736011). ZW
acknowledges grant from the Centre for Quantum Technologies, and the
WBS grant under contract no. R-710-000-008-271 and R-710-000-007-271.
BZ is supported by NSERC Discovery Grant 400500 and CIFAR.

%


\end{document}